\documentclass{elsart3p}
\usepackage{graphicx}

\usepackage{amssymb}

\begin{document}

\begin{frontmatter}


\title{Bright solitary waves of trapped atomic Bose-Einstein condensates}
\author{N. G. Parker\thanksref{label1}}\ead{n.p@physics.org}, \author{A. M. Martin\thanksref{label2}}, \author{C. S. Adams \thanksref{label3}} and \author{S. L. Cornish\thanksref{label3}}

\address[label1]{Department of Physics and Astronomy, McMaster University, Hamilton, Canada}
\address[label2]{School of Physics, University of Melbourne, Parkville,
Victoria 3010, Australia}
\address[label3]{Department of Physics, Durham University, Durham, DH1 3LE, UK}

\begin{abstract}
Motivated by recent experimental observations, we study theoretically multiple bright solitary waves of trapped Bose-Einstein condensates.  Through variational and numerical analyses, we determine the threshold for collapse of these states.  Under $\pi$-phase differences between adjacent waves, we show that the experimental states lie consistently at the threshold for collapse, where the corresponding in-phase states are highly unstable.   Following the observation of two long-lived solitary waves in a trap, we perform detailed three-dimensional simulations which confirm that in-phase waves undergo collapse while a $\pi$-phase difference preserves the long-lived dynamics and gives excellent quantitative agreement with experiment.  Furthermore, intermediate phase differences lead to the growth of population asymmetries between the waves, which ultimately triggers collapse.  
\end{abstract}


\end{frontmatter}

% main text
\section{Introduction}

The nonlinearity present in atomic Bose-Einstein condensates has led
to demonstrations of fascinating self-trapped states known as
solitons. These one-dimensional wavepackets, well known in
non-linear optics and other fields \cite{soliton_book},
arise when the nonlinearity of the medium counter-acts the effects
of dispersion.  They have been realized in several distinct
matter-wave forms: bright \cite{khaykovich,strecker,cornish}, dark
\cite{dark} and gap \cite{gap} solitons.  The bright
solitons manifest themselves as self-trapped lumps of matter, held
together by attractive atomic interactions.  In three-dimensions, bright matter-wave ``solitons'' are self-trapped
in one dimension and require external confinement in the remaining
two directions.  We will henceforth refer to these 3D solitonic
states as {\em bright solitary waves} (BSWs).  Due to their
self-trapping properties, matter-wave BSWs offer significant
possibilities in atom-optical applications such as atom
interferometry \cite{strecker} and probing the atom-surface
interaction \cite{physica}. However, the three-dimensional nature
of BSWs leads to the presence of an undesirable collapse instability
when the attractive interactions become too strong
\cite{perez_garcia,carr_castin,salasnich,parker_JPB}.  Not only does
this affect the static properties of BSWs, it can also
destabilise their collisions
\cite{carr,khaykovich2,phase_diff,parker_JPB2}. Although techniques
to suppress collapse effects in attractive BECs have been proposed,
e.g., by applying rotation about the {\it z}-axis \cite{anis} or by
time-modulating the scattering length \cite{feshbach_management},
the collapse instability ultimately remains.  A detailed
understanding of the properties of matter-wave BSWs and their
regimes of collapse is therefore essential for the advancement of
this field.

The presence of solitonic solutions is revealed by considering the
mean-field and zero-temperature limit of atomic BECs. Here the BEC
`wavefunction' $\psi({\bf r},t)$ satisfies a nonlinear wave equation
known as the Gross-Pitaevskii equation (GPE) \cite{dalfovo}, given
by,
\begin{equation}
i\hbar \frac{\partial \psi}{\partial
t}=\left[-\frac{\hbar^2}{2m}\nabla^2 + V_{\rm ext}({\bf r})
+\frac{4\pi\hbar^2 a_{\rm s}}{m}|\psi|^2 \right]\psi,
\end{equation}
where $m$ is the atomic mass.   The nonlinear term arises from the
short-range atomic interactions, characterised by the {\it s}-wave
scattering length $a_{\rm s}$, and can be repulsive ($a_{\rm s}>0$)
or attractive ($a_{\rm s}<0$). In 1D and in the absence of an
external potential $V_{\rm ext}({\bf r})=0$, this has an identical form
to the 1D cubic nonlinear Schr{\"o}dinger equation and supports the
exact dark and bright soliton solutions derived by Zakharov and
Shabat \cite{shabat}.  Key properties of these 1D solitons are that
they can have any population, their collisions are elastic
\cite{gordon} and they are stable to thermal dissipation.  However, atomic BECs are intrinsically 3D objects, typically confined by
harmonic traps of the form $V_{\rm ext}({\bf r})=m\omega_r^2(r^2
+\lambda^2 z^2)/2$, where $\omega_r$ and $\lambda \omega_r$ are the
radial and axial trap frequencies, respectively.  The presence of
the extra dimensions modifies the special properties of the
1D soliton by introducing a critical atomic population, inelastic
collisions \cite{parker_JPB2} and thermal dissipation \cite{sinha}.

The critical population and inelastic collisions of BSWs arise due
to the collapse instability which affects attractively-interacting
BECs in general.  It is convenient to introduce the dimensionless
interaction parameter $k$ defined as,
\begin{equation}
k=\frac{N|a_{\rm s}|}{a_r},
\end{equation}
where $N$ is the number of atoms in the system and $a_r=\sqrt{\hbar/m\omega_r}$ is the radial harmonic oscillator length of the trap \footnote{Note that, in some other studies \cite{ruprecht,gammal,yukalov,roberts}, this quantity is defined using the geometric mean of the harmonic oscillator lengths $a_{\rm 3D}=\sqrt{\hbar/m(\omega_r^2\omega_z)^{1/3}}$ rather than the radial quantity $a_r$.}.  Collapse of the system occurs when $k$ exceeds a critical value $k_{\rm c}$, which is typically of the order of unity.  This implies that there is a critical population $N_{\rm c}$ beyond which collapse occurs.  Note that it is the presence of trapping that leads to a finite value of $N_{\rm c}$, whereas a homogeneous, untrapped BEC is always unstable to collapse \cite{nozieres}.

In a recent experiment multiple BSWs were generated in a three-dimensional trap \cite{cornish}.  Intriguingly, the ensuing dynamics were remarkably robust and long-lived.  In this work we analyse the properties of bright solitary matter-waves under external confining potentials with particular emphasis on this experiment  \cite{cornish}.  After reviewing the details of this experiment (Section 2), we employ a variational approach and full solution of the Gross-Pitaevskii equation to study the ground state and first-excited state solutions of the system (Section 3).  We extend this to additional solitary waves using a dynamical model, revealing the threshold of the collapse instability for up to four solitary waves and compare to experimental measurements.  We then directly simulate the experimental oscillations of two BSWs (Section 4) and show that a $\pi$-phase difference is essential to support the observed dynamics and gives excellent agreement with the experimental data.  Finally, we present the conclusions of our work (Section 5).

%The GP equation has been demonstrated to be an
%excellent model of condensate dynamics at the mean-field,
%weakly-interacting level \cite{dalfovo}. Although the GP equation is
%insufficient to explicitly describe collapse dynamics, where higher
%order effects such as three-body loss and heating become
%considerable \cite{milstein}, it still provides a good model to
%infer the onset of collapse instabilities \cite{ruprecht,gammal}.

\section{The JILA experiment}
\label{sec:expt} We will review the BSW experiment of Cornish {\it
et al.} \cite{cornish} and its observations. Firstly, a stable
$^{85}$Rb BEC was formed with repulsive {\it s}-wave interactions.
This typically contained $15 000$ atoms with less than $500$ thermal
atoms. The magnetic confining trap was cylindrically symmetric with
radial frequency $\omega_r=2\pi \times 17.3$Hz and is weakly
elongated with a trap ratio $\lambda=0.4$.  For this system,
experimental \cite{roberts} and theoretical work
\cite{parker_JPB,gammal} agrees that the critical interaction
parameter for collapse is $k_{\rm c} \approx 0.64$. Following
previous experiments \cite{roberts}, the {\it s}-wave scattering
length was then quickly tuned to be attractive by means of a
molecular Feshbach resonance \cite{Feshbach}. At this point the number of atoms in
the system greatly exceeded the critical number triggering a collapse.
During the collapse, three-body
atomic losses rise and eventually stabilised the condensate.  The
``remnant" condensate typically contained more than the critical number of atoms, $N_{\rm c}$, and was clearly divided in the axial direction into a symmetric arrangement of distinct
wavepackets, i.e. bright solitary waves, that oscillated along the weaker axial direction of confining potential maintaining a stationary centre of mass. Note that it is thought
that the collapsing condensate fragments into multiple wavepackets
via a modulational instability
\cite{carr,salasnich_mod_inst}.  Up to six BSWs were observed,
depending on factors such as the final scattering length and the initial
number of atoms \cite{cornish}.
\begin{figure}[t]
\centering
\includegraphics[clip=true]{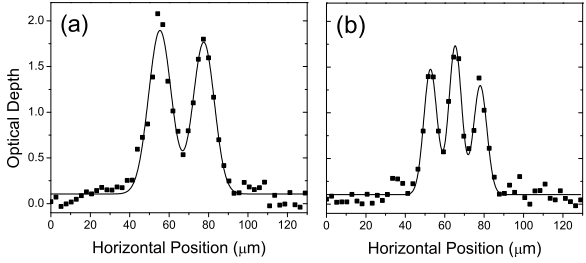}
\caption{Axial profile of the optical depth from the JILA experiment
following BSW formation showing (a) two BSWs and (b) three BSWs. The profiles were taken when the BSWs reached the outer turning points of their oscillatory motion in the harmonic trap.  In
each case, the profile is fitted with the form $\sum_{i=1}^{N_{\rm BSW}}\left[\alpha_i \exp\left\{-(z-\beta_i)^2/\sigma_i^2\right\}\right]$, where $N_{\rm BSW}$ is the number of BSWs and $\alpha_i$, $\beta_i$ and $\sigma_i$ are fitting parameters. In (a), the mean BSW width is $\overline{\sigma}\approx 7.8(4)\mu$m and in (b) it is $\overline{\sigma}\approx 4.5(4)\mu$m.} \label{fig:experiment}
\end{figure}
In Fig.~\ref{fig:experiment} we present the typical appearance of the
condensate column density following the formation of the BSWs.  The images were taken when the BSWs reached the outer turning points of their oscillatory motion in the harmonic trap and show a
cross-section of the optical depth in the axial
direction for the case of two and three BSWs.  Note that the
density profiles are approximately symmetric about the origin.
In particular, quantitative measurements of the experimental system
were made for fixed scattering length $a_{\rm s}=-0.6$nm and
approximately $4000$ atoms. Since the system initially contained
approximately $500$ thermal atoms, we will assume the number of
condensate atoms to be $N=3500$.  This corresponds to an interaction
parameter of $k=0.8$ which exceeds the critical interaction
parameter $k_{\rm c}$.  Despite this, the observed dynamics were
surprisingly stable with negligible dissipation over $3$~s. For the case of  two BSWs, the dynamics consist of oscillations in anti-phase along the axial
direction of the trap, with the BSWs colliding repeatedly at the trap centre. It
is thought that the stable dynamics were supported by the existence
of a repulsive $\pi$-phase difference between the BSWs, which
allowed each BSW to support an atom number corresponding to an interaction parameter just below $k_{\rm c}$. Indeed, experimental
measurements for up to four BSWs showed that the the average value
of $k$ per BSW never exceeded $k_{\rm c}$ and this has been recently found to be in good agreement with a theoretical study \cite{parker_JPB}.   Note that the observed BSW dynamics showed no significant thermal dissipation, despite the presence of the highly energetic burst of atoms ejected from the condensate during the collapse \cite{donley}. It is likely that these excitations are so ``hot'' ($T_{\rm burst}\sim 50\,\mbox{nK}$) and dilute that thermal equilibrium is not reached over the experimental timescales, with the ``thermal atoms'' remaining effectively invisible to the BSWs throughout.  Note
that the existence of a $\pi$-phase difference has also been inferred in
the experiment of Strecker {\it et al.} \cite{strecker} by the
observed repulsive interactions between the BSWs.

\section{BSW solutions and the collapse instability}

In 3D and in the presence of interactions there are no exact
analytic solutions of the GPE and solutions must be obtained
numerically or via approximated approaches. In the latter case, variational methods \cite{perez_garcia,carr_castin,salasnich,parker_JPB} provide considerable insight without the the need for full numerical solutions.  Here we will consider both a variational approach and numerical solution of the GPE.

\subsection{Ground state solutions}

Firstly, we shall study the ground state solutions in the system. We
approximate the ground state solutions (denoted by the
$0$-subscript) by a single-peaked ansatz of the form
\cite{perez_garcia,carr_castin,salasnich,parker_JPB},
\begin{equation}
\psi_0(r,z)=\sqrt{\frac{N}{2\pi L_z L_r^2}}\exp\left(
-\frac{z^2}{2L_z^2}\right)\exp \left( -\frac{r^2}{2L_r^2} \right).
\label{eqn:ansatz}
\end{equation}
where $L_z$ and $L_r$ represent the axial and radial sizes,
respectively.  The energy of the system $\varepsilon$ is defined by
the GP energy functional via,
\begin{eqnarray}
\varepsilon=\int d^3 {\bf r} \left[\frac{\hbar^2}{2m}|\nabla
\psi|^2+V_{\rm ext}|\psi|^2+\frac{2\pi\hbar^2 a_{\rm s}}{m}|\psi|^4
\right]. \label{eqn:E_func}
\end{eqnarray}
By substituting the ansatz into the energy functional of
Eq.~(\ref{eqn:E_func}) we determine the variational energy for the
ground state $\varepsilon_0$. For convenience we employ the rescaled
parameters $l_r=L_r/a_r$, $l_z=L_z/a_r$ and $E=\varepsilon/(N\hbar
\omega_r)$, where $a_r=\sqrt{\hbar/m\omega_r}$ is the radial
harmonic oscillator length. Furthermore, we introduce our interaction parameter $k=N|a_{\rm s}|/a_r$.  The
ground state ansatz energy then becomes,
\begin{equation}
E_0=\frac{1}{2}\left(\frac{1}{l_r^2}+\frac{1}{2l_z^2}\right)+
\frac{1}{2}\left(l_r^2+\frac{\lambda^2l_z^2}{2}\right)-\frac{k}{\sqrt{2\pi}l_z
l_r^2}. \label{eqn:E_land1}
\end{equation}
The first group of terms represents kinetic energy, the second group
represents potential energy and the final term represents the energy
arising from the {\it s}-wave interatomic interactions. This
equation defines an energy landscape for the system in terms of
$l_z$ and $l_r$, and variational solutions correspond to local
energy minima in this landscape.  We denote the widths of the
variational solution by $l_z^0$ and $l_r^0$.  The variational method
has been employed successfully to study the ground states of the
system \cite{perez_garcia,parker_JPB,carr}. In particular, in the
absence of axial trapping $(\lambda=0)$, local energy minima arise
which correspond to self-trapped BSW solutions. When $k$ exceeds a
critical value the local energy minimum ceases to exist and the
global energy minimum, which occurs at the origin, dominates the
system.  This corresponds to collapse.

\begin{figure}[t]
\centering
\includegraphics[width=7cm,clip=true]{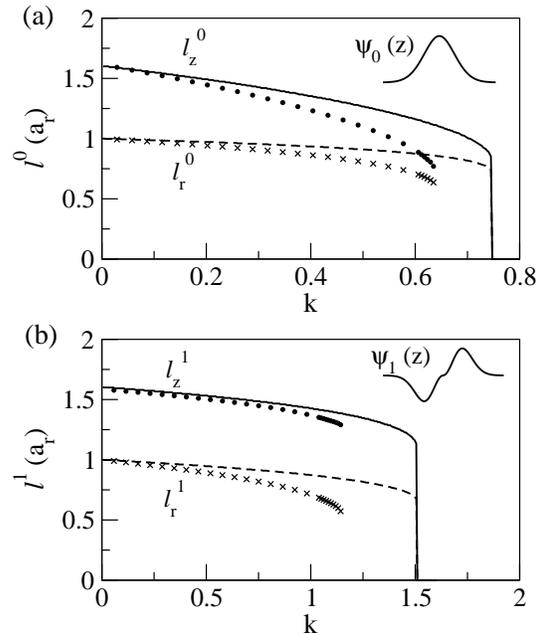}
\caption{(a) Ground state solutions under axial trapping
$\lambda=0.4$ according to the variational approach of
Eq.~(\ref{eqn:E_land1}) (solid/dashed lines) and by numerical solution of
the GPE (crosses/circles).  Upper and lower lines indicate the axial
lengthscale $l^0_z$ and radial lengthscale $l^0_r$ of the solution,
respectively. (b) First-excited state solutions for $\lambda=0.4$
according to the variational energy of Eq.~(\ref{eqn:E_land2}) and
the full numerical solution of the GPE.  } \label{fig:lengthscales}
\end{figure}

We have calculated the variational solutions under an axial trap
defined by $\lambda=0.4$.  The size of the variational solutions as
a function of interaction parameter $k$ are presented in Fig.~\ref{fig:lengthscales}(a).  For $k=0$, the solution corresponds to
the exact non-interacting gaussian ground state with
$l_z=\sqrt{\hbar/m\omega_z}=\lambda^{-1/2}a_r=1.58a_r$.  As the
attractive interactions grow in size, the solutions, which are always
elongated in the $z-$direction, shrink in both dimensions.  Finally,
at $k=0.746$, the variational solutions disappear and the system is
unstable to collapse.

We have also calculated the exact solutions by solving the GPE
numerically.  This is performed using the imaginary-time propagation
technique: under the substitution $t\rightarrow -it$ in the GPE, the
equation evolves to the ground state of the system, providing it
exists\footnote{Note that the equation is no longer unitary and so
must be renormalised at each time step to preserve atom number}. The
axial and radial lengthscales of the GPE solutions correspond to the
distance over which the density decreases by a factor $1/e$ from its
peak value.  The GPE lengthscales are presented in
Fig.~\ref{fig:lengthscales}(a) by crosses (filled circles) for the radial (axial) direction. For $k=0$ the GPE results
agree exactly with the variational prediction. As $k$ is increased
the GPE predictions decrease at a faster rate than the variational
method. Furthermore, the GPE solutions become unstable to collapse
for $k_{\rm c}=0.630 (5)$. This is approximately $15\%$ lower than
the variational method and arises because the variational method
consistently over-estimates the BSW widths and therefore under-estimates
the peak density, which is the trigger of collapse.  This result is
consistent with previous studies \cite{parker_JPB,gammal}, where
$k_{\rm c}$ was mapped out for a range of trap ratios.
In a previous JILA experiment \cite{roberts}, the critical
interaction parameter for collapse (of a single wavepacket) was
measured to be $k_{\rm c}=0.64 (7)$ \footnote{Note that the quoted
value of $k_{\rm c}$ in \cite{roberts} was revised in
\cite{claussen} based on more accurate measurements of $a_{\rm
s}$.}. This is in excellent agreement with the GPE prediction of
$k_{\rm c}=0.630 (5)$, as noted elsewhere \cite{parker_JPB,gammal}.

Comparison of the experimental BSW density profiles (e.g. Fig.~\ref{fig:experiment}) to the theoretical predictions is, however, unsuccessful.  According to the theoretical results of Fig.~\ref{fig:lengthscales}(a), the axial width of the BSW should be at the very most equal to the non-interacting value of $l_z^0(k=0)=1.6a_r\approx 4\mu$m. However, the mean fitted width of each BSW is approximately $7.8\mu$m in Fig.~\ref{fig:experiment}(a) and $4.5\mu$m in Fig.~\ref{fig:experiment}(b).  This discrepancy is almost certainly due to the low resolution of the experimental imaging \cite{cornish}.

\subsection{First-excited state solutions}
Since two (and more) BSWs were also observed in the JILA experiment,
and are believed to be supported by a $\pi$-phase difference, it is
pertinent to consider the first-excited state of the system (denoted by $1$-subscript).  We
extend the variational approach by replacing the gaussian axial
profile with that of the first-excited harmonic oscillator state,
such that the ansatz is,
\begin{eqnarray}
\psi_1(r,z)&=&\sqrt{\frac{2N}{\pi^{3/2}L_z
L_r^2}}\left(\frac{z}{L_z}\right) \nonumber
\\
&\times&\exp \left(-\frac{z^2}{2L_z^2}\right)\exp \left(
-\frac{r^2}{2L_r^2} \right), \label{eqn:ansatz2}
\end{eqnarray}
The density profile of this ansatz is double-peaked and can be
interpreted as two BSWs featuring a $\pi$-phase difference. Note that
a similar approach by Michinel {\it et al.} \cite{michinel} employed
Hermite functions to study multiple BSWs in a trap.

Following the same method as for the ground state ansatz, we arrive
at the variational energy for the first-excited state,
\begin{eqnarray}
E_1&=&\frac{1}{2}\left(\frac{1}{l_r^2}+\frac{3}{2l_z^2}\right)+
\frac{1}{2}\left(l_r^2+\frac{3\pi^2\lambda^2l_z^2}{2}\right)
\nonumber
\\
&-&\frac{3k}{4\sqrt{2\pi}l_z l_r^2}. \label{eqn:E_land2}
\end{eqnarray}
We have obtained the corresponding first-excited variational
solutions and plotted their lengthscales in
Fig.~\ref{fig:lengthscales}(b).  Collapse of the first-excited
variational solution occurs at $k_{\rm c}=1.508$.
We have also obtained the exact first-excited states of the full
GPE, shown in Fig.~\ref{fig:lengthscales}(b) by crosses. This was
performed using the imaginary-time technique while enforcing the
wavefunction to be asymmetric via $\psi(z)=-\psi(-z)$. These
solutions become unstable at $k_c=1.145(5)$.
Note that in the absence of axial confinement, there is no
stationary first-excited state, because the BSWs exert a repulsive force on each other which decays exponentially with their separation.  The lowest energy state is therefore when the BSWs are infinitely separated.

According to both the variational ansatz and the full GPE solution,
the first-excited state supports an interaction parameter which is
almost twice that of the ground state.  Note that if the individual
BSWs were completely independent we would expect the system to
support exactly twice $k_{\rm c}$ of the individual BSWs. 

\subsection{Up to four BSWs}

\begin{figure}[t]
\centering
\includegraphics[width=6.5cm,clip=true]{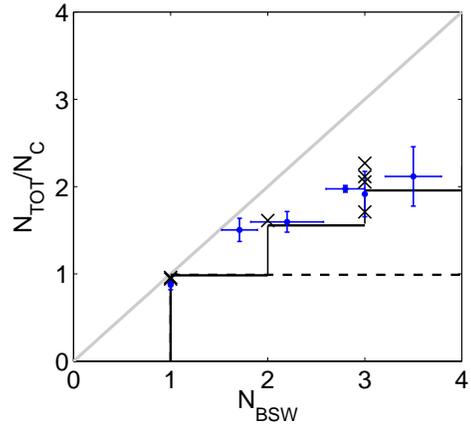}
\caption{The ratio $N_{\rm tot}/N_{\rm c}$ as a function of the number of BSWs $N_{\rm BSW}$.  GPE simulations show the critical points for phases differences of $\Delta \phi=0$ (dashed line) and $\Delta \phi=\pi$ (bold solid line), for a fixed scattering length of $a_{\rm s}=-0.6$nm.  Above/below these lines the configurations are stable/unstable to collapse.  The experimental JILA data, shown by points with error bars, is obtained at various scattering lengths.  The corresponding GPE results are presented by black crosses, where it is assumed that $10\%$ of the experimentally detected atoms were non-condensed.  The function $N_{\rm tot}=N_{\rm BSW}N_{\rm c}$ is shown for comparison (grey line).}
\label{fig:Ns}
\end{figure}

In the JILA experiment, quantitative measurements were made of up to four BSWs, with the results presented in Fig.~\ref{fig:Ns} (points with error bars).  While the total atom number typically exceeded the critical atom number for the ground state $N_{\rm c}$, the {\em average} atom number per BSW was less than, or approximately equal to, $N_{\rm c}$.  This observation is thought to be a direct consequence of repulsive $\pi$-phase differences between the BSWs.
We now consider configurations of up to four BSWs.  In principle one could extend the variational approach to model any number of BSWs by using higher-order excited states of the harmonic oscillator.  However, here we will employ numerical solutions of the full GPE.  States of one and two BSWs are obtained by imaginary time propagation as detailed above.  States of three and four BSWs cannot be formed by imaginary time propagation since the lowest energy state, which is where the atoms populate the central BSWs, is unstable to collapse.  A dynamical method is employed \cite{parker_JPB} where we begin with a ground state repulsively-interacting BEC.  The interactions are then switched to the required attractive value while simultaneously imposing a periodic distribution of $\pi$-phase steps. In this manner dynamic states of three or four BSWs are created and their critical regime for collapse can be probed.  The results for a fixed scattering length of $a_{\rm s}=-0.6$~nm are shown by the solid line in Fig.~\ref{fig:Ns}.  Above (below) this line, the configurations are unstable (stable).  We see that four BSWs are readily supported, providing $\pi$-phase differences are present.  For $0$-phase differences, only one BSW with up to $N_{\rm c}$ can be supported, as indicated  by the dashed line.   Assuming the BSWs to be independent and that each contains up to $N_{\rm c}$ atoms, the function $N_{\rm tot}=N_{\rm BSW}N_{\rm c}$ (grey line) is satisfied.  The numerical results deviate from this function as $N_{\rm BSW}$ increases.  This is due to the presence of interactions between the BSWs and the unequal distribution of atoms in the BSWs, i.e. the central (outer) BSWs contain more (less) atoms, as observed in the experiment.

The experimental number measurements (points with error bars) were obtained at various scattering lengths.  Using these scattering lengths, we have numerically evaluated the critical points for collapse according to the GPE.  Note that we have assumed $10\%$ of the experimental atom number to be non-condensed.  The experimental data is in excellent agreement with these GPE predictions, with every data point showing consistency between theory and experiment.  These results show the importance of $\pi$-phase differences. Furthermore, they show that the experimental system consistently forms a state which is right at the limit of collapse.  This is a remarkable effect, given that the system is initially in a state which is highly unstable to collapse.

\section{Dynamics of two BSWs in a trap}

The collision of two BSWs in a homogeneous waveguide has been
considered previously \cite{khaykovich2,phase_diff,parker_JPB2}.
During the collision, a high density state forms during the collapse
and providing the interaction parameter is sufficiently large, this
can induce collapse. The collisions are most prone to collapse when
(i) the BSWs are in-phase, since this maximises the overlap and hence the peak density during the
collision, and (ii) for low speed collisions, since the timescale over
which this overlap occurs is large.  In contrast, the collapse
instability is heavily suppressed when the BSWs feature a
$\pi$-phase difference or if the impact speed is high, since this reduces the timescale over which a collapse can
occur. Furthermore, we have recently shown that when the relative
phase $\Delta \phi$ between the colliding BSWs does not equal zero
or $\pi$, a sizeable population transfer can occur between the BSWs
\cite{parker_JPB2}. For $0<\Delta \phi<\pi$, this population
transfer flows in one direction, while for $\pi<\Delta \phi<2\pi$,
it flows in the opposite direction.  While these studies have
typically considered single collisions in a homogeneous waveguide,
the presence of axial trapping enables multiple collisions.
One-dimensional approaches to BSW collisions in a trap have been
made, with a recent study employing a particle model for the BSWs
\cite{martin}.  Below we make a detailed study of the 3D dynamics of two BSWs under the conditions of the JILA experiment.

In the experiment, the axial
density profile was measured at regular intervals and each profile was fitted to a single gaussian profile to give the axial full-width-half-maximum (FWHM). The experimental data showing the evolution of the FWHM is
presented in Fig.~\ref{fig:collisions}(iii).  The key observations
are that the FWHM oscillates primarily at $2\lambda \omega_r$ and
that there is negligible dissipation over $3$~s which corresponds to $40$ oscillations. As discussed in Section \ref{sec:expt}, the condensate contains approximately $3500$ atoms.  We assume each BSW to contain $N=1750$
atoms and therefore $k \approx 0.4$. At the extreme points of their
motion, the BSWs were observed to be displaced by approximately
$z_0=16\mu$m from the trap centre.  Note this this relatively large separation means that it is more appropriate here to consider the system as a collection of single BSWs rather than a true first-excited state of the system.  
Our initial state therefore
consists of two ground state BSWs with $k=0.4$, positioned at
$z_0=\pm 16\mu$m.  In addition, we also impose a relative phase
difference $\Delta \phi$ between the initial BSWs. Typically, the
experimental BSW density profile remained symmetric about the trap
centre, as illustrated in the experimental profiles shown in
Fig.~\ref{fig:experiment}. Since only relative phase differences of
$0$ and $\pi$ (modulo $2\pi$) preserve this symmetry in BSW
collisions \cite{parker_JPB2} we will concentrate on these phase
differences.

Note that the corresponding dynamics for three BSWs in a trap appear significantly less stable than the two-BSW state, even with $\pi$-phase differences.  This will be considered in a future study.

\subsection{Relative phases $\Delta \phi=0$ and $\pi$}
\begin{figure}[t]
\centering
\includegraphics[width=7.5cm,clip=true]{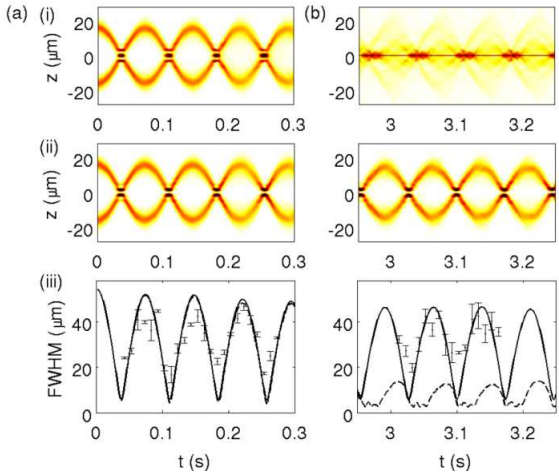}
\caption{(a) Evolution of the radially-integrated axial density
at early times for two BSWs ($N_{\rm S}=1750$, $z_0=\pm 16\mu$m,
$\omega_z/2\pi=6.8$~Hz and $\omega_r/2\pi=17.3$~Hz) with (i) $\Delta
\phi=0$ and (ii) $\Delta \phi=\pi$. (iii) Full-width-half-maximum FWHM
of a gaussian fit to the axial density for $\Delta \phi=0$ (dashed
line), $\Delta \phi=\pi$ (solid line), and experimental data
(points) \cite{cornish}. Note that in order to match the phase of
the oscillations we have shifted the experimental data in time.
(b) Same as (a) but for late times.} \label{fig:collisions}
\end{figure}

Simulation of the dynamics for short times are shown in
Fig.~\ref{fig:collisions}(a) for (i) $\Delta\phi=0$ and (ii) $\Delta
\phi=\pi$. Initially the BSWs accelerate identically towards the
trap centre. As they interact the relative phase becomes apparent
with the BSWs overlapping for $\Delta \phi=0$ and `bouncing' for
$\Delta \phi=\pi$. However, apart from at the point of interaction,
the BSW dynamics are almost identical over this time scale. The
corresponding FWHM is presented in Fig.~\ref{fig:collisions}(a)(iii)
and is in excellent agreement with the experimental data.  The
$2\lambda \omega_r$ oscillations in the FWHM arises since each BSW
oscillates in the axial direction at a frequency of $\lambda \omega_r$.
Note that while the simulated FWHM begins at a maximum, the
experimental data begins at a minimum \cite{cornish} since the BSWs
are created in close proximity. Consequently, we have shifted the
experimental data in Fig.~\ref{fig:collisions} by a quarter of the
oscillation period such that the experimental and simulated data are
in phase.  Also note that due to the resolution limit in the
experiment, the FWHM cannot be resolved below approximately
$15\mu$m.

The corresponding dynamics for late times are shown in
Fig.~\ref{fig:collisions}(b).  The $\Delta \phi=0$ BSWs have fully
collapsed, no longer matching the experimental results. In contrast,
the $\Delta \phi=\pi$ collisions remains practically unaffected.
Furthermore, they are in excellent agreement with the experimental
data. This shows that a phase difference of $\pi$ is crucial to
support the long-lived oscillations observed in the JILA experiment.
The presence of $\pi$-phase differences was also inferred in the BSW
experiment of Strecker {\it et al.} \cite{strecker}.

\subsection{Intermediate relative phases $0<\Delta \phi<\pi$}
\begin{figure}[t]
\centering
\includegraphics[width=7.5cm,clip=true]{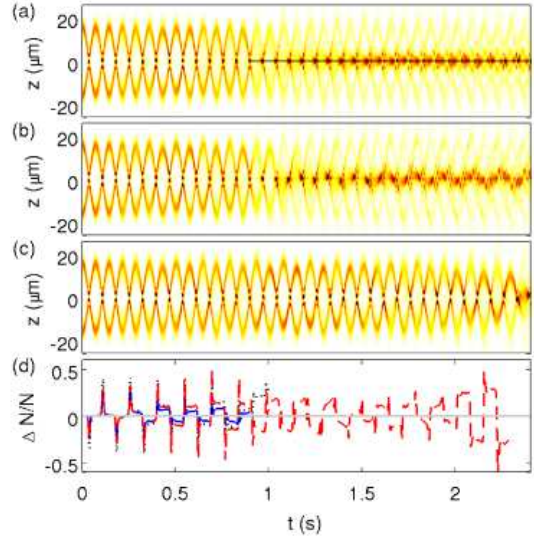}
\caption{Dynamics of two BSWs ($k=0.4$) in the $\lambda=0.4$ JILA
trap with (a) $\Delta \phi=\pi/4$, (b) $\Delta \phi=\pi/2$ and (c)
$\Delta \phi=3\pi/4$.  The initial separation is $z_0=16\mu$m. (d)
Population difference across the system $(N_{\rm L}-N_{\rm R})/N$
for $\Delta \phi=\pi/4$ (solid blue line), $\phi=\pi/2$ (dotted
black line) and $3\pi/4$ (dashed red line).  For the cases of
$\Delta \phi=0$ and $\pi$, the population difference is zero at all
times, as indicated by the grey horizontal line.}
\label{fig:collisions3}
\end{figure}
For comparison, we have also simulated the dynamics of the two BSWs
for intermediate relative phases in the range $0<\Delta \phi<\pi$.
The density dynamics are presented in Fig.~\ref{fig:collisions3} for
(a) $\Delta \phi=\pi/4$, (b) $\pi/2$ and (c) $3\pi/4$. Additionally,
in Fig.~\ref{fig:collisions3}(d) we plot the population difference
about the trap origin $\Delta N/N=(N_{\rm L}-N_{\rm R})/N$, where
$N_{\rm L}$ is the atom number for $z<0$, $N_{\rm R}$ is the atom
number for $z>0$ and $N$ is the initial atom number in each BSW.

Although the BSWs start with equal populations in each case, an
asymmetry develops over time with one BSW becoming increasingly
populated at the expense of the other. This is due to a
population transfer during each collision, as observed in
\cite{parker_JPB2} for a single BSW collision.  Here the BSWs collide with approximate speed $v=\lambda \omega_r z_0\approx 0.7{\rm mm~s}^{-1}$ at the trap centre.  At this relatively large speed, the population transfer is small and, from \cite{parker_JPB2}, we can expect it to be of the order of $1\% N$.  However, the multiplicity of the collisions leads to the growth of a noticeable population transfer, until eventually the system becomes unstable against collapse. We note that the time at which the system becomes unstable depends upon the initial phase. During each collision, the relative phase is also modified and after many collisions this can lead to a reversal of the population transfer.  For example, for $\Delta \phi=3\pi/4$, the population difference oscillates slowly before the system ultimately collapses.  Since no large asymmetries are observed in the experimental density
profiles, this validates our original assumption that the relative
phase must of either zero or $\pi$ (or very close).
Furthermore, the fact that all cases except $\pi$ become unstable in
well under $3$~s gives further evidence that a $\pi$-phase difference
is crucial to maintain the dynamics of the two BSWs observed in the experiment.

\section{Conclusions}

In summary, we have performed a detailed analysis of the multiple bright solitary waves (BSW) observed experimentally \cite{cornish}. We confirm that such multiple BSW states are only stable if there is a $\pi$ phase difference between each BSW. This allows each BSW to contain approximately the critical number of atoms.  Remarkably the experimental data implies that the atom number in each BSW lies consistently just under the threshold for collapse.  We find that two BSWs featuring a $\pi$-phase difference undergo stable dynamics in a trap over long times (of order a few seconds or over 40 collisions), in excellent quantitative agreement with experimental measurements.  In contrast, for $0$-phase difference, the system undergoes collapse.  For intermediate relative phases ($0<\Delta \phi<\pi$) we observe significant population transfer between the BSWs and a lifetime against collapse that depends upon the value of the relative phase. We propose that these predictions could be verified by using a phase imprinting technique similar to that used in the creation of dark solitons \cite{dark} to impose a controlled relative phase on the BSWs shortly after their creation.

\ack We thank S. A. Gardiner and D. H. J. O'Dell for many stimulating discussions. We acknowledge support from the Canadian Commonwealth Scholarship program (NGP), ARC (AMM), UK EPSRC (CSA/SLC) and the Royal
Society (SLC).

\end{document}